\documentclass[font=modern,mono=courier]{linkup}

\usepackage{xspace}
\usepackage[T1]{fontenc}
\usepackage{XCharter}
\usepackage[xcharter,vvarbb]{newtxmath}
\usepackage[scaled=0.92]{helvet}
\usepackage{enumitem}
\usepackage{booktabs}
\usepackage{adjustbox}
\usepackage{amsmath,amsfonts}
\usepackage[noend]{algorithmic}
\usepackage[ruled]{algorithm2e}
\usepackage{array}
\usepackage{textcomp}
\usepackage{xcolor}
\usepackage{microtype} 
\definecolor{linkup-green}{HTML}{EB653F}
\usepackage{stfloats}

\usepackage{url}
\usepackage{verbatim}
\usepackage{graphicx}
\usepackage{xcolor}
\usepackage{xspace}
\usepackage[acronym]{glossaries}
\usepackage{bbm}
\usepackage{multirow}
\usepackage{tikz}
\usepackage{hyperref}
\usepackage{bbm}
\RequirePackage{tikz}
\RequirePackage{nicematrix}
\providecolor{linkup-green}{HTML}{F26B1D}

\setcitestyle{numbers,square,comma}
\title[SPARSEUP: Towards Better Recipes for Training Modern Learned Sparse Retrievers]{SPARSEUP: Towards Better Recipes for Training Modern Learned Sparse Retrievers}
\reporttypelabel{Technical Report}
\author[]{Thibault Formal}
\author[]{Pierre Erbacher}
\author[]{Cosimo Rulli}
\author[]{Paulo Roberto Moura Junior}
\affiliation[]{Linkup, Paris}

\DeclareRobustCommand{\nameicon}{%
  \raisebox{-0.2em}{\includegraphics[height=1.5em]{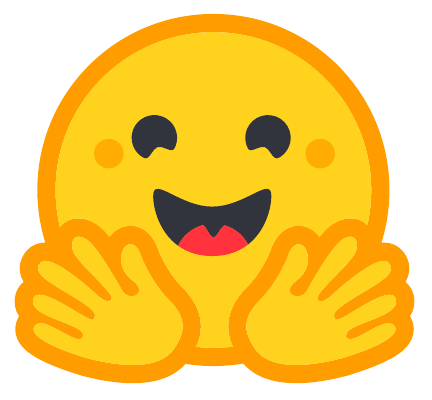}}%
}
\abstract{%
Learned sparse embeddings are a powerful alternative to dense retrieval, yet they have fallen behind single-vector and late interaction models. One reason for their limited adoption is that they are notoriously difficult to train, generally requiring very tight conditions to converge. We present a recipe for training \splade-like encoders on modern backbones and data that achieves a better efficiency-effectiveness trade-off. We discuss the role of the logit distribution and its impact on training stability and the final sparsity of the encoder. We also introduce a new per-position fan-out control --- a hard cap on how many vocabulary dimensions each position may emit. Finally, we merge redundant vocabulary entries, such as case and word-boundary
variants of the same word, which are common in cased BPE-based backbones. Together, these three ingredients let us train \sparseup, a new family of
sparse encoders built on \modernbert and fine-tuned on \lateon's data. We then scale \sparseup with large-scale contrastive pre-training, distillation, and Matryoshka-style training. The resulting \sparseupvtwo reaches 57.2 average nDCG@10 on \beir-15, matching \lateon, the best public retriever under 150M parameters we know of, while being an order of magnitude more efficient.
}

\definecolor{lightgreen}{HTML}{7FBD9C}
\definecolor{linkup-green-dark}{HTML}{3D6650}

\hypersetup{
  colorlinks=true,
  linkcolor=lightgreen,   
  urlcolor=black,    
  citecolor=lightgreen    
}

\definecolor{linkup-green}{HTML}{7FBD9C}
\newlength{\lknumwd}
\newcommand{\lknumband}[3]{%
  \settowidth{\lknumwd}{\textbf{00.0}}%
  \tikz \fill[linkup-green!25, rounded corners=3pt]
        ([xshift=-\tabcolsep-\lknumwd-3pt]#1-|\inteval{#3+1}) rectangle
        ([xshift=-\tabcolsep+3pt]\inteval{#2+1}-|\inteval{#3+1});}
\newcommand{\lkcolbest}[2]{%
  \settowidth{\lknumwd}{\textbf{00.0}}%
  \tikz \fill[linkup-green!70, rounded corners=2pt]
        ([xshift=-\tabcolsep-\lknumwd-2pt,yshift=-1pt]#1-|\inteval{#2+1}) rectangle
        ([xshift=-\tabcolsep+2pt,yshift=1pt]\inteval{#1+1}-|\inteval{#2+1});}

\metadata[\nameicon]{\texttt{\href{https://huggingface.co/Linkup-Platform/linkup-sparseup-embed-v1}{$\blacksquare$ Linkup-Platform/linkup-sparseup-embed-v1}}}
\metadata[]{\texttt{\href{https://huggingface.co/Linkup-Platform/linkup-sparseup-embed-v2-pretrain}{$\blacksquare$ Linkup-Platform/linkup-sparseup-embed-v2-pretrain}}}
\metadata[]{\texttt{\href{https://huggingface.co/Linkup-Platform/linkup-sparseup-embed-v2}{$\blacksquare$ Linkup-Platform/linkup-sparseup-embed-v2}}}
\metadata[Playground]{\texttt{\href{https://huggingface.co/spaces/Linkup-Platform/sparseup-embed-v2-visualization}{$\blacksquare$ Linkup-Platform/sparseup-embed-v2-visualization}}}
\date{}
\usepackage{soul}
\usepackage{xcolor}

\newcommand{\msmarco}{\textsc{Ms Marco v1}\xspace}

\newcommand{\beir}{\textsc{Beir}\xspace}
\newcommand{\decbeir}{\textsc{Decon-Beir}\xspace}

\newcommand{\seismic}{\textsc{Seismic}\xspace}

\newcommand{\splade}{\textsc{Splade}\xspace}
\newcommand{\lm}{\textsc{LM}\xspace}
\newcommand{\spladeT}{\textsc{Splade}-v3\xspace}

\newcommand{\laconic}{\textsc{Laconic}\xspace}
\newcommand{\lateon}{\textsc{LateOn}\xspace}
\newcommand{\denseon}{\textsc{DenseOn}\xspace}

\newcommand{\modernbert}{\textsc{ModernBERT}\xspace}
\newcommand{\sparseup}{\textsc{SPARSEUP}\xspace}
\newcommand{\sparseupvtwo}{\textsc{SPARSEUP}-v2\xspace}

\newcommand{\data}{\textsc{Embeddings-Fine-Tuning}\xspace}
\newcommand{\ettin}{\textsc{Ettin}\xspace}

\newcommand{\roberta}{\textsc{RoBERTa}\xspace}

\newacronym{nns}{\textsc{Nns}}{Nearest Neighbor Search}
\newacronym{anns}{\textsc{Anns}}{Approximate Nearest Neighbor Search}
\newacronym{smips}{Sparse \textsc{Mips}}{Sparse Maximum Inner Product Search}
\newacronym{ts}{\textsc{TS}}{Threshold Sampling}
\newacronym{knn}{\textsc{$\kappa$-nn} graph}{$\kappa$-Nearest Neighbor graph}
\newacronym{sketch}{$\alpha$-MSS}{$\alpha$-Mass Subvector Sketch}
\newacronym{setsketch}{Set $\alpha$-MSS}{Set $\alpha$-Mass Subvector Sketch}
\newacronym{nir}{\textsc{Nir}}{Neural Information Retrieval}
\newacronym[plural=\textsc{Sae}s,firstplural=Sparse Autoencoders (\textsc{Sae}s)]{sae}{\textsc{Sae}}{Sparse Autoencoder}
\newacronym[plural=\textsc{Llm}s,firstplural=Large Language Models (\textsc{Llm}s)]{llm}{\textsc{Llm}}{Large Language Model}

\begin{document}
\maketitle


\section{Introduction} \label{sec:introduction}

Learned sparse retrieval offers an effective alternative to single-vector dense retrieval. Beyond its efficiency advantages~\citep{bruch2024seismic}, it forms a distinct class of models with their own strengths and weaknesses, such as a better ability to match rare words precisely~\citep{10.1007/978-3-030-99739-7_14}. The recent success of late interaction models~\cite{sourty2026denseonlateonfullyopen,krimmel2026pplxembedv2latetechnicalreport} has renewed interest in sparse representations. Because late-interaction pipelines are inherently multi-stage, they require efficient first-stage candidate generation, and several recent methods perform this step using sparse vectors and inverted indices~\cite{guo2026kmeanssinglestagesparsecoding,10.1145/3805712.3809920,10.1145/3511808.3557325}.

The \splade series~\citep{splade-sigir2021,formal2021splade,lassance2024spladev3} set the sparse standard for years, but follow-up work has been relatively scarce~\citep{10.1145/3726302.3730192,awasthy2025graniteembeddingmodels,10.1145/3805712.3809565,formal2026learning,moura2026slideon}. While dense and late interaction encoders keep benefiting from newer backbones and data~\citep{sourty2026denseonlateonfullyopen,zhang-etal-2024-mgte,vera2025embeddinggemmapowerfullightweighttext,eslami-etal-2026-diffusion}, learned sparse models have been slower to follow: they do not transfer seamlessly to modern encoders~\citep{DBLP:conf/sigir/GengY26,jang2026rescaling}, and few recent works have really pushed them forward --- with LLM-based approaches~\citep{xu-etal-2025-csplade,xu2026laconicdenseleveleffectivenessscalable,10.1145/3726302.3730225} being the main exception. Two recent papers trace the poor behavior of modern encoders in sparse retrieval to different causes. Geng and Yang~\cite{DBLP:conf/sigir/GengY26} mainly point to the vocabulary: modern byte-level BPE tokenizers keep case and word-boundary variants as separate entries, so a single term is spread over several surface forms~\citep{10.1007/978-3-032-21289-4_33}. They introduce a Vocabulary Transfer technique which migrates any backbone to the \texttt{bert-base-uncased} vocabulary by re-initializing the \lm head and adapting the embeddings with a light MLM phase. In contrast, Jang \emph{et al}.~\cite{jang2026rescaling} attribute instability to scale: large-norm heads (e.g., \modernbert~\citep{warner-etal-2025-smarter}) inflate the dot product, and rescaling the head at initialization restores effectiveness. Moreover, sparse retrievers are generally harder to train, and their reliance on a tokenizer vocabulary as the output space makes them less suited to multilingual~\citep{nguyen2026milco,formal2026learning} or multimodal retrieval~\citep{cho2026inferencefreemultimodallearnedsparse}. These limitations have gradually pushed them out of the spotlight.

In this paper, we build on top of previous work by providing an effective recipe to train sparse text encoders using modern model backbones and data. We first study the logit distribution of different models, showing how a right-shifted logit distribution negatively affects training stability and effectiveness. We show that \emph{logit shifting} --- introduced as APC by Geng and Yang \cite{DBLP:conf/sigir/GengY26} --- is a simple and effective way to address this problem across backbones. We further introduce \emph{fan-out}, a sparsity-inducing mechanism that limits the number of vocabulary terms each input position can activate before pooling, showing that this approach outperforms existing post-hoc pruning or regularization-based approaches --- in addition to stabilizing training. Modern backbones typically rely on byte-level BPE tokenizers, whose vocabularies contain several surface forms of the same term. We therefore apply \emph{vocabulary folding}, which merges these near-duplicates in the encoder output, consistently during training and inference. 

We apply this recipe to \modernbert~\citep{warner-etal-2025-smarter}, \ettin~\citep{weller2026seq}, and \roberta~\citep{liu2019robertarobustlyoptimizedbert}, turning backbones that underperform, or even fail to train, under the standard \splade recipe into effective sparse encoders. Among our contributions, \emph{logit shifting} and \emph{fan-out control} each improve sparse encoder quality, and their gains are complementary. Trained on fine-tuning data alone, the resulting \modernbert-based model, \sparseup, already achieves strong effectiveness. We then integrate the recipe into a more complete training pipeline that combines contrastive pre-training on large-scale weakly supervised data, cross-encoder distillation, and Matryoshka representation learning. The resulting \sparseupvtwo reaches \textbf{57.2 nDCG@10 on \beir-15}, outperforming \denseon~\citep{sourty2026denseonlateonfullyopen} and \textsc{SlideOn}~\citep{moura2026slideon} and matching \lateon, which are, to our knowledge, the strongest publicly available encoders under 150M parameters. Since these models are trained on closely comparable data and pipelines, this constitutes a controlled comparison across dense, late-interaction, and sparse paradigms. \sparseupvtwo also generalizes well beyond \beir: it achieves the best nDCG@10 on CURE-en, Finance, and BRIGHT-Pro, with large margins over dense and late-interaction models on the latter two (e.g., $+13.3$ over \lateon on BRIGHT-Pro). Across the full benchmark suite, it outperforms \denseon on 10 and \textsc{SlideOn} on 11 of 13 benchmarks, while \lateon remains stronger on long-document retrieval (MLDR) and on LIMIT. Out of domain, sparse representations thus sit between dense and late-interaction models in effectiveness, consistent with prior findings~\citep{10.1145/3634912,10.1145/3726302.3730225}, while being substantially cheaper than the latter.
\section{Methodology} \label{sec:methodology}
\smallskip
\noindent \textbf{Notation}. 
A \splade encoder maps the backbone's hidden states $H \in \mathbb{R}^{s \times d}$ into the vocabulary $\mathcal{V}$ with the Language Modeling head (\lm) --- obtaining the logits $L = \mathtt{LM}(H) \in \mathbb{R}^{s \times |\mathcal{V}|}$. The output weight $w_j$ of the $j$-th vocabulary term is then obtained by max-pooling saturated activations over the input positions:
\begin{equation}
    w_j = \max_{i \in \{1,\dots,s\}} \log\left(1 + \mathrm{ReLU}(L_{ij})\right), \quad j \in \{1,\dots,|\mathcal{V}|\}.
    \label{eq:splade}
\end{equation}

Sparsity in \splade vectors is usually obtained through  regularization~\citep{Paria2020Minimizing,10.1145/3726302.3730192,10.1145/3726302.3730163} or post-hoc pruning~\citep{lassance2023static-pruning}, which leaves practitioners to balance effectiveness against efficiency. In the following, we introduce three simple ingredients that act on the model itself and improve this trade-off for modern backbones --- in addition to greatly stabilizing training.
\subsection{Logit Shifting --- Better Initialization}
\label{subsec:logshift}
\begin{figure}[!tb]
    \centering
    \includegraphics[width=\textwidth]{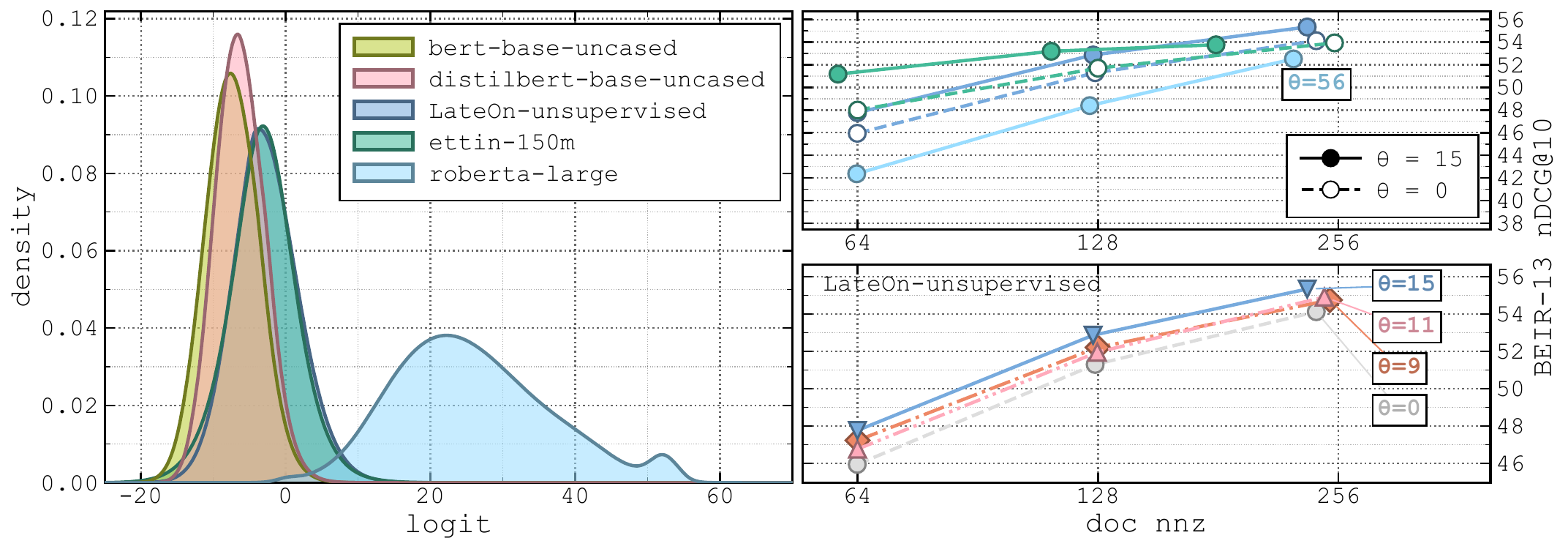}
    \caption{{\it (Left)} Distribution of the \lm head logits computed on 200$k$ \msmarco passages. {\it (Right)} Logit Shift ablation for multiple backbones and multiple shift values. {\it Note that the \roberta encoder does not train without the shift}.}
    \label{fig:hist_logits}
\end{figure}

Recall from Equation~\ref{eq:splade} that a vocabulary term gets a non-zero weight as soon as one of its $s$ logits is positive, so the more positive entries $L$ has, the denser the initial representation. Pre-training does not control this: the softmax cross-entropy is invariant to a global additive offset, so the absolute scale of the logits is arbitrary. \splade-like encoders, however, apply the saturating activation $\log(1 + \mathrm{ReLU}(z))$, which is sensitive to exactly that scale. The \splade head thus inherits an operating point that pre-training never set. We hypothesize that the fraction of positive logits at initialization (as well as their magnitude) is a proxy for how ``ready'' a backbone is to be trained as a sparse encoder.

Figure~\ref{fig:hist_logits} {\it (Left)} shows the logit distributions of \textsc{Bert-Base}, \textsc{Distilbert-Base} (both uncased), \roberta-Large, and two modern backbones: the pre-trained \modernbert variant used by \textsc{LateOn} (\textsc{LateOn-unsupervised}) and \ettin-150M. The two \textsc{BERT}-based models on which \splade has been successfully trained~\citep{splade-sigir2021,formal2021splade,lassance2024spladev3} are the most left-shifted: fewer logits are positive, so the initial representation is already partly sparse. \modernbert and \ettin are shifted further right, consistent with their reported suboptimal training~\citep{DBLP:conf/sigir/GengY26,warner-etal-2025-smarter}, which our experiments confirm (Figure~\ref{fig:hist_logits}, {\it Right, top}, dotted lines). \roberta-Large is the most right-shifted and is notoriously hard to train as a sparse encoder.

Two effects explain this. First, when most logits are positive, the representation is dense and activations sit where the slope of $\log(1 + \mathrm{ReLU}(z))$ is small: values are compressed into a narrow range and the gradient signal that discriminates between terms weakens. Second, the sparsity regularizer must prune thousands of terms while competing with the ranking loss, whereas a sparse start leaves far fewer terms to remove.

Geng and Yang~\cite{DBLP:conf/sigir/GengY26} propose Vocabulary Transfer (VT), a multi-component recipe whose most effective part is Activation Potential Calibration (APC). APC shifts the logits by a constant $\theta$ before the activation. This sparsifies the representation without any extra pre-training and moves the logits back to the high-slope region, where gradients are informative and less noisy. In Section~\ref{sec::ablation}, we provide a better understanding of the role of $\theta$ and derive a rule of thumb for setting it on backbones that previously underperformed or even failed to train, namely \modernbert, \ettin, and \roberta-Large (Figure~\ref{fig:hist_logits}, {\it Right, top}). 

\begin{equation}
    w_j = \max_{i \in \{1,\dots,s\}} \log\left(1 + \mathrm{ReLU}(L_{ij} - \theta)\right).
    \label{eq:shift}
\end{equation}

\subsection{Fan-out --- Limiting Expansion}

We additionally introduce a complementary approach that limits the capacity of expansion. In Equation~\ref{eq:splade}, every input position may activate any of the $|\mathcal{V}|$ vocabulary dimensions. We argue that an input term is adequately represented by a small set of related terms, and that expansions beyond this neighborhood contribute to noise rather than signal. Unconstrained expansion also burdens the optimization, which must learn to suppress, at every position, a long tail of irrelevant terms over the entire vocabulary. We therefore restrict the expansion of each position to its $f$ highest-scoring vocabulary terms before pooling. The weight of the $j$-th term becomes:
\begin{equation}
    w_j = \max_{i \in \{1,\dots,s\}} \mathbbm{1}\left[j \in \mathcal{T}_f(i)\right] \log\left(1 + \mathrm{ReLU}\left(L_{ij} - \theta\right)\right),
    \label{eq:fanout}
\end{equation}
where $\mathcal{T}_f(i) $ are the $f$ vocabulary terms with the largest logits at position $i$. Note that per-position fan-out control is related to, but not equivalent to, enforcing a fixed budget directly by applying a top-$k$ operator to the pooled \splade vector instead of using regularization. The latter approach generally underperforms at budgets other than the one used during training (Figure~\ref{fig:prune_and_fanout}, {\it left}), although Matryoshka-style losses can mitigate this and yield good trade-offs across budgets (Section~\ref{sec:sparseup-v2}). Fan-out control also resembles the per-position top-$k$ activations used in SAE-based sparse encoders~\cite{formal2026learning,10.1145/3805712.3809565,moura2026slideon,clavié2026latenttermsdenseretrievers}.

\subsection{Vocabulary Folding --- Simple Bandage for Modern Tokenizers}

Most modern encoders rely on byte-level BPE tokenizers, which are case-sensitive and distinguish word-initial tokens, prefixed with the space marker \texttt{Ġ}, from word-internal ones. A single word thus maps to up to four vocabulary entries, e.g., \texttt{paris}, \texttt{Paris}, \texttt{Ġparis}, and \texttt{ĠParis}. This redundancy is implicit for dense or late interaction models, but it directly hinders sparse retrieval, which uses tokens as embedding dimensions: a query term only matches a document term with the exact same surface form, so the model must either spread the weight of each term over all its variants or accept lexical mismatches. Geng and Yang~\cite{DBLP:conf/sigir/GengY26} identify this \emph{vocabulary gap} as one of the main reasons why modern encoders lag behind BERT in sparse retrieval, and close it by migrating the backbone to a normalized vocabulary via embedding initialization and MLM adaptation. We apply a simpler alternative that acts only on the output space. Surface variants are grouped by their lowercased body with the word-start or continuation marker removed. The canonical representative is an existing vocab id: the lowercased space-marked form when it exists (\texttt{Ġparis}), otherwise the bare lowercase form, etc. Let $\pi: \mathcal{V} \rightarrow \mathcal{V}'$ send each term to that representative, with $\mathcal{V}'\subset\mathcal{V}$. After pooling, each canonical term takes the largest weight among its surface forms, and every other coordinate is set to zero:
\begin{equation}
    w'_u = \max_{t:\pi(t)=u} w_t.
    \label{eq:folding}
\end{equation}

Folding is parameter-free, and requires no additional training --- we apply it at both training and inference time. For \modernbert, it reduces the vocabulary size from $|\mathcal{V}| \approx 50k$ to $|\mathcal{V}'| \approx 34k$ while improving the effectiveness--efficiency trade-off, yielding a gain of about 1.5 points in average nDCG@10 on \beir-13 together with sparser representations (not shown).

\begin{figure}[tb]
    \centering
    \includegraphics[width=1.0\linewidth]{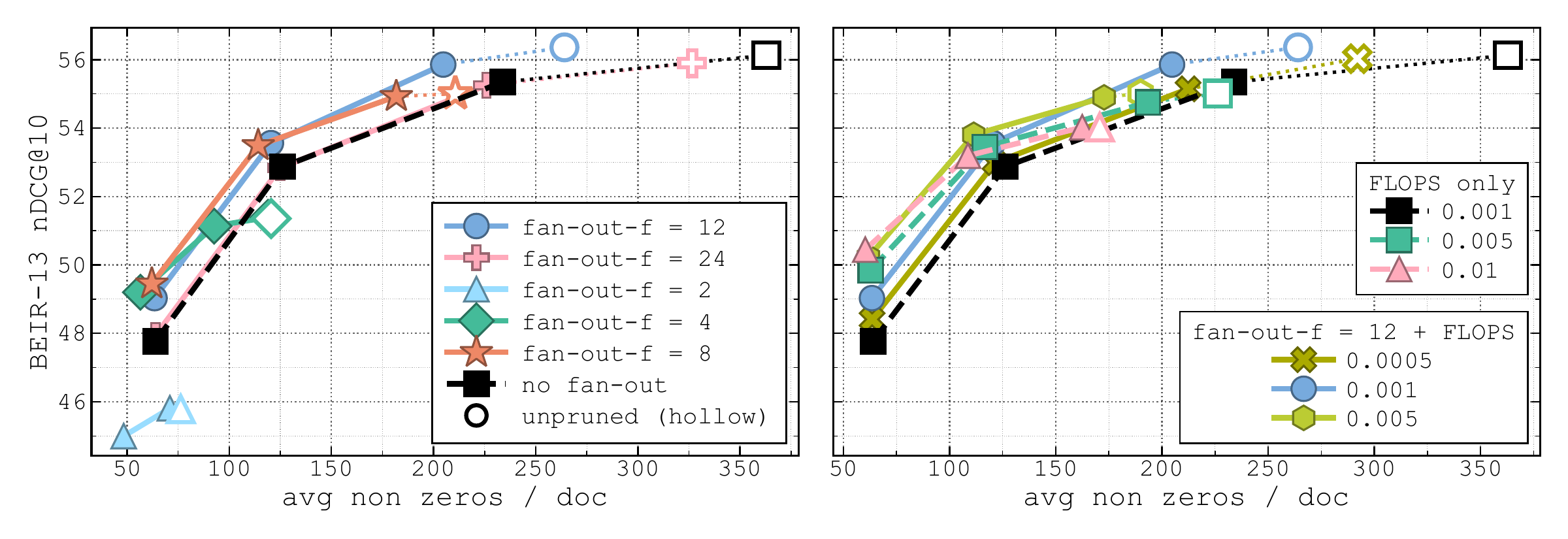}
    \caption{{\it (Left)} Fan-out ablation --- nDCG@10 on \beir-13 as a function of non-zero values per document. {\it (Right)} Regularization grid search; overall, fan-out control improves effectiveness at most sparsity levels. $\theta=15$ for all the runs.}
    \label{fig:fanoutabl}
\end{figure}

\section{SPARSEUP --- Design Choices that Matter}\label{sec:experiments}

We study \sparseup in two stages. First, we run controlled ablations: starting from a contrastively pre-trained checkpoint, we fine-tune models while varying one design choice at a time (\Cref{sec::ablation}). Second, we bring these findings into a full modern training recipe, \sparseupvtwo
(\Cref{sec:sparseup-v2}), and measure the added effect of large-scale contrastive pre-training, distillation, and Matryoshka-style training. Together, these changes yield a sparse retriever that matches strong late-interaction models at roughly an order of magnitude less compute.


\begin{figure}
    \centering
    \includegraphics[width=1.0\linewidth]{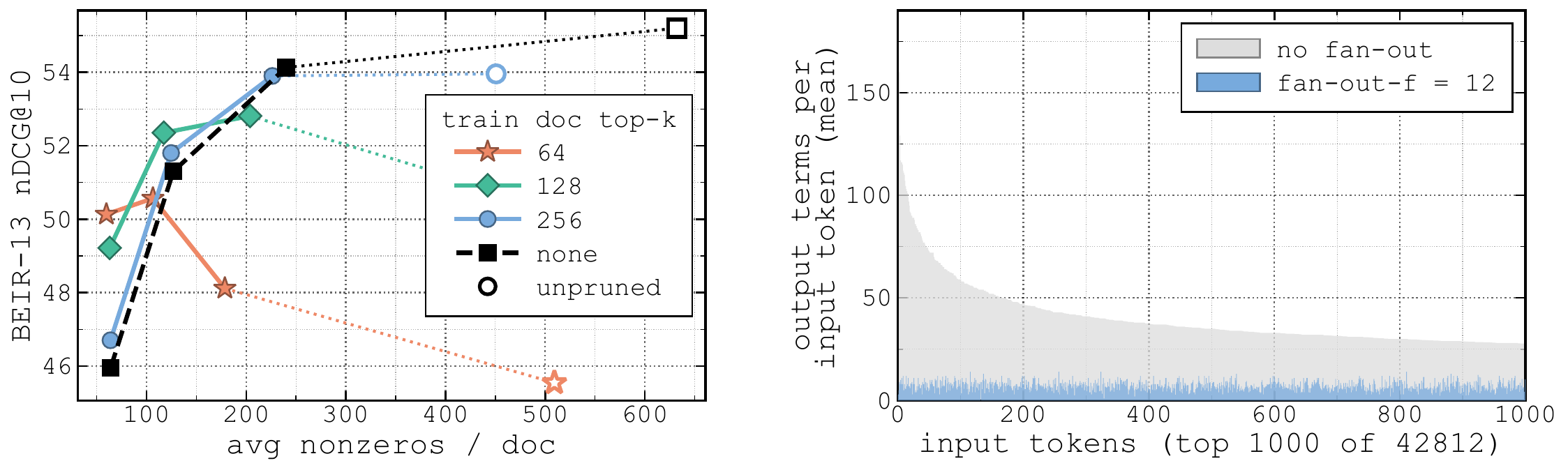}
    \caption{\textit{(Left)} Pruning ablation --- Models are trained at specific document pruning values $\{64, 128, 256\}$ and evaluated for various pruning. \textit{(Right)} Average number of output terms per input token on a sample of $50k$ documents, without (grey) and with (blue) \textit{fan-out}.}
    \label{fig:prune_and_fanout}
\end{figure}

\subsection{Experimental Setting} 

\smallskip 
\noindent \textbf{Training}. We fine-tune all models on the \data dataset,\footnote{\href{https://huggingface.co/datasets/lightonai/embeddings-fine-tuning}{\textsc{lightonai/embeddings-fine-tuning}}} mirroring the training setup of \denseon and \lateon: InfoNCE with $\tau{=}6$,\footnote{Unlike dense retrievers, sparse encoders require a larger temperature.} seven sampled hard negatives per query, and in-batch negatives. We otherwise follow the standard \splade setup, using the FLOPS regularizer~\citep{Paria2020Minimizing} with $\lambda_q = \lambda_d = 10^{-3}$ by default.

\smallskip
\noindent \textbf{Initialization}. The ablations in Section~\ref{sec::ablation} focus on fine-tuning. Accordingly, \sparseup is initialized from a strong contrastively pre-trained late-interaction checkpoint, the one used to train \lateon\footnote{\href{https://huggingface.co/lightonai/LateOn-unsupervised}{\textsc{lightonai/LateOn-unsupervised}}}, onto which we graft \modernbert's MLM head. For \sparseupvtwo (Section~\ref{sec:sparseup-v2}), we extend the pipeline with our own large-scale sparse contrastive pre-training\footnote{\href{https://huggingface.co/datasets/lightonai/embeddings-pre-training}{\textsc{lightonai/embeddings-pre-training}}} starting from \modernbert, cross-encoder distillation, and Matryoshka-style training, and show that this yields stronger results.

\smallskip
\noindent \textbf{Evaluation \& Fair Efficiency Comparison}. We primarily report nDCG@10 on \beir~\citep{thakur2021beir}, and provide additional out-of-domain results in Table~\ref{tab:offshelf}. Because sparse models can reach different effectiveness levels simply by activating more terms, comparing them at their natural sparsity would favor the densest one. We therefore evaluate every model at several document pruning budgets $\text{nnz}_d \in \{64, 128, 256\}$, as well as without pruning. Queries are capped at $64$. Retrieval is performed with \seismic~\citep{bruch2024seismic,Bruch2024SeismicWave,ecir-seismic,nardini2026compact} using default hyperparameters. Our evaluation setting places post-hoc pruning, varying regularization strengths, direct training at a target pruning level, and our recipe on a common effectiveness–efficiency front. {\bf \texttt As we show below, the \sparseup recipe achieves the best trade-off overall}. All results in the tables are reported with unpruned representations, whose sparsity remains comparable to that of standard learned sparse models: for instance, \sparseupvtwo has on average 40 and 180 non-zero entries per query and per document, respectively, on the MS MARCO passage collection.

\subsection{Ablation Study}\label{sec::ablation}

In this section, we discuss the role of the components introduced in Section~\ref{sec:methodology} and how they affect the performance of \sparseup.

\smallskip
\noindent\textbf{Logit Shift}. 
We complement the analysis of Section~\ref{subsec:logshift} by discussing the choice of $\theta$. Empirically, setting $\theta$ around the $99.9$$th$ percentile of the logit distribution --- which can be estimated cheaply on a small sample of the training collection --- gives consistent results across backbones.  Figure~\ref{fig:hist_logits} {\it (Right, bottom)} reports the effectiveness for different values of $\theta$ for \textsc{LateOn-unsupervised}: values close to the 99.9$th$ percentile ($13.9$) perform comparably --- we thus report $\theta=15$. When $\theta$ is too high, models collapse during training (not enough positive values).  When $\theta$ is too low, effectiveness degrades, as the trade-off becomes harder to control. \ettin has a very similar distribution to \modernbert, so we adopt $\theta = 15$ as well. As \roberta has a more right-shifted distribution, we set $\theta=56$. 

\smallskip
\noindent\textbf{Fan-out}. 
Figure~\ref{fig:fanoutabl} {\it (Left)} shows the impact of fan-out on the effectiveness--sparsity trade-off. We train the \lateon backbone with $\text{fan-out-}f \in \{2,4,8,12,24\}$ and compare it against the same model trained without fan-out --- keeping all other settings fixed. 
Intermediate values ($f \in \{8,12\}$) dominate the no-fan-out baseline, and we use $f = 12$ for \sparseup. A natural question is whether the same frontier could be obtained simply by increasing the sparsity pressure. Figure~\ref{fig:fanoutabl} {\it (Right)} compares the $f    = 12$ model with models trained without fan-out and with stronger FLOPS regularization. Regularization alone moves models towards sparser representations, but it does so at a higher cost in effectiveness, and its curves remain below that of fan-out at comparable sparsity. This matches our observation during training that sparsifying a model through the regularizer alone is harder to control.

\smallskip
\noindent\textbf{Top-$k$ Pruning During Training}.
In the experiments above, models were trained without any constraint on the number of non-zero entries in their representations. In practice, however, representations are typically pruned to their top-$k$ entries at inference time to reduce memory and compute costs. We therefore investigate whether enforcing a fixed number of non-zero values per document during training improves performance under inference-time pruning. Figure~\ref{fig:prune_and_fanout} compares models trained with top-$k \in \{64, 128, 256\}$ against an unconstrained baseline. Under strong compression ($k = 64$), training with top-$k$ pruning (pink) yields a clear benefit, improving nDCG@10 on \beir-13 by 4 points over the unconstrained model (black). This advantage gradually vanishes as $k$ increases. Moreover, models trained with a given top-$k$ do not generalize to larger values of $k$ at inference, where their performance degrades substantially. Without inference-time pruning, the unconstrained model remains the best overall. These controlled comparisons show that combining logit shifting and fan-out control with weak regularization yields the best effectiveness--efficiency trade-off. In Section~\ref{sec:sparseup-v2}, we further extend this setting with Matryoshka-style training, which pushes the Pareto front even further for \sparseupvtwo.

\begin{figure}[t]
  \centering
  \includegraphics[width=\linewidth]{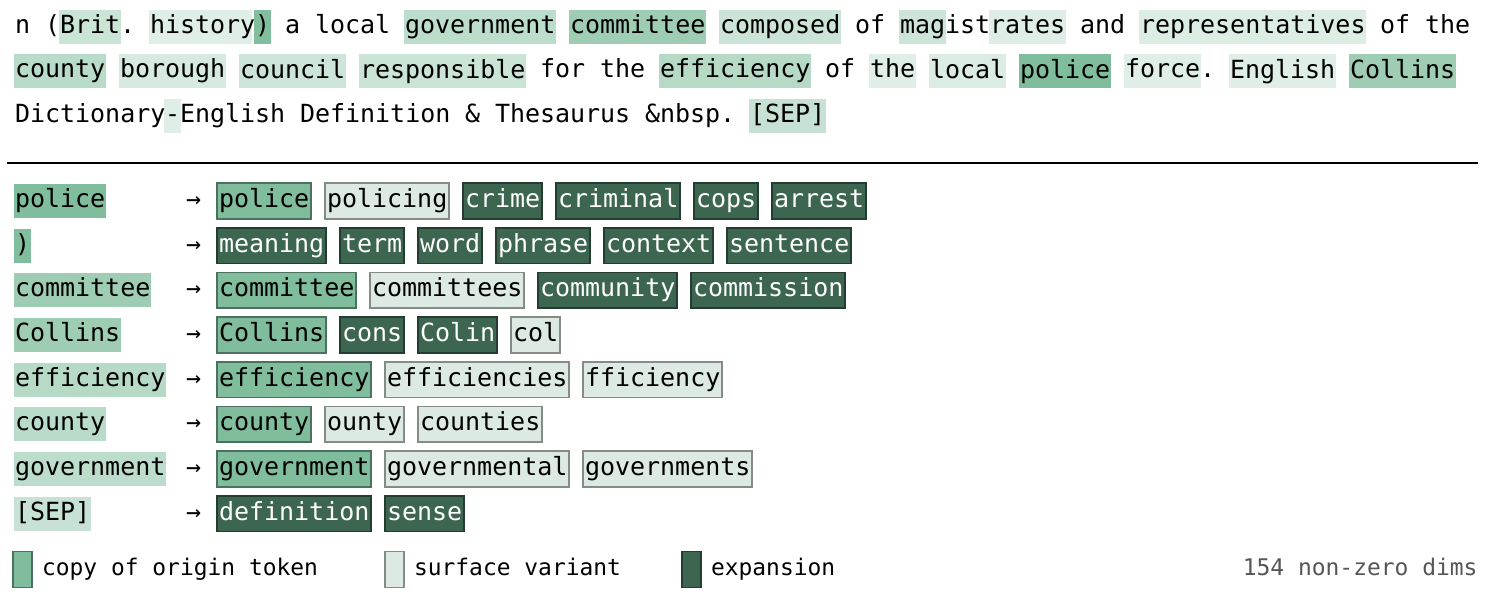}\\[4pt]
  \vspace{1.0cm}
  \includegraphics[width=\linewidth]{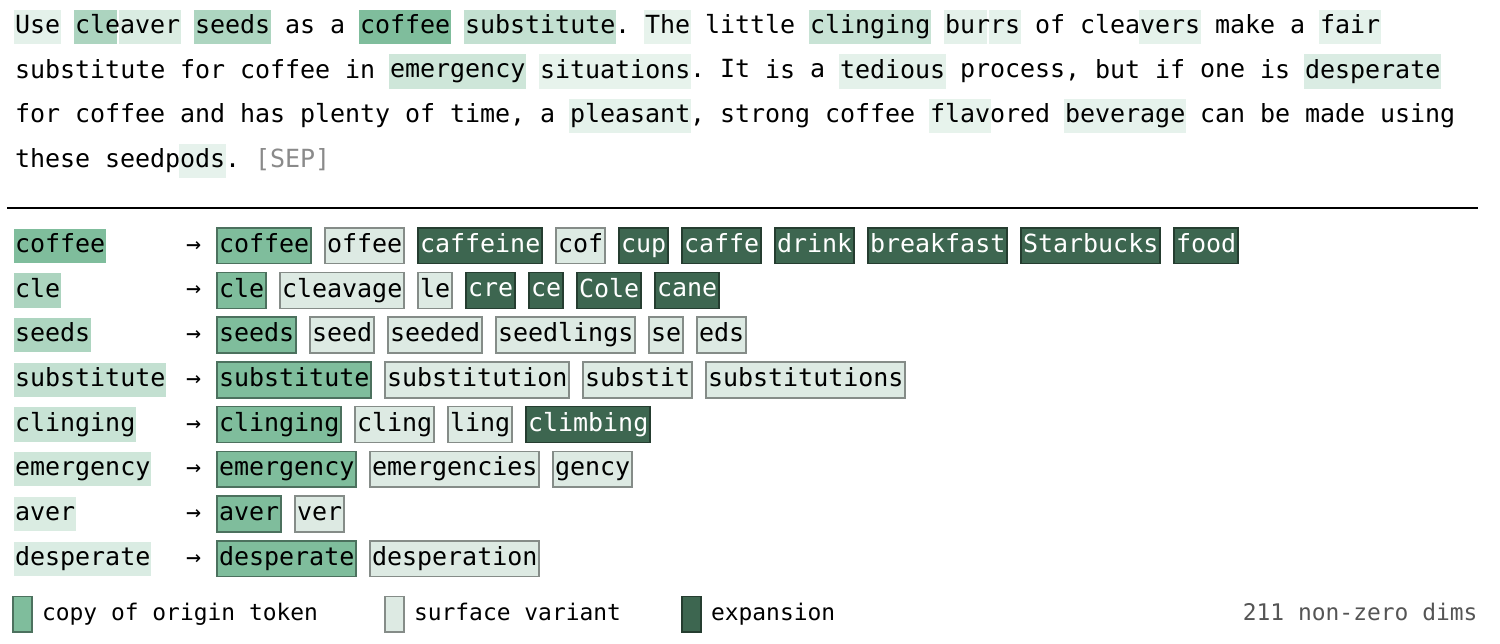}\vspace{1.0cm}
  \includegraphics[width=\linewidth]{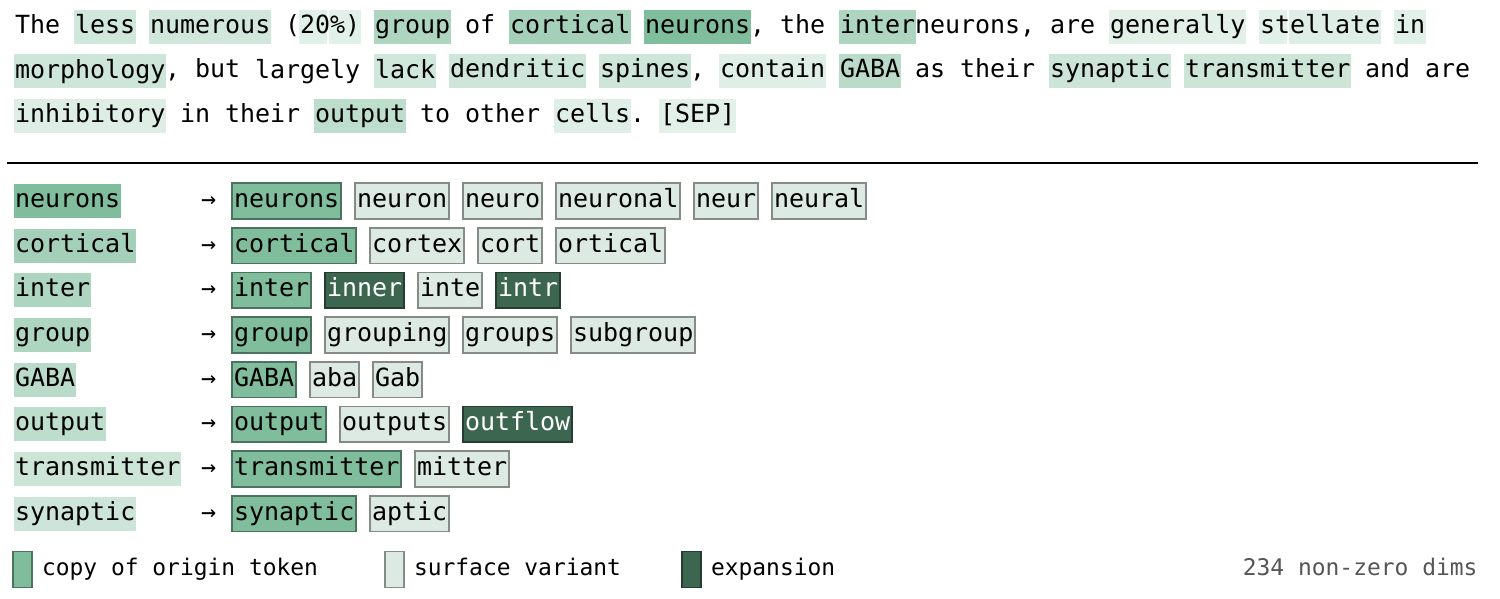}\\[4pt]
  \caption{\sparseup document representations. {\it (Top)} input tokens. {\it (Bottom)} Terms emitted by the 8 strongest positions, colored by whether the term copies its origin token, is a surface variant of it, or is a new expansion.}
  \label{fig:sparse-examples}
\end{figure}

\noindent\textbf{Qualitative Examples}. 
Figure~\ref{fig:sparse-examples} shows some qualitative examples for \sparseup. The model exhibits the term re-weighting, removal, and expansion behavior now typical of \splade-like models. However, because of the BPE tokenizer, words that are split into several sub-tokens tend to activate irrelevant sub-pieces, e.g., \texttt{cle}$\rightarrow\{$\texttt{cre}, \texttt{ce}, \texttt{Cole}, \texttt{cane}$\}$. These are the so-called \emph{wacky weights}~\citep{mackenzie2021wacky}, which are more pronounced on the query side, where shorter inputs provide less context. The effect persists even though fan-out partly mitigates it. Thus, while the representations look qualitatively less clean than those of models such as \spladeT, the model remains substantially more effective.

\begin{table}[t]
\centering
\caption{nDCG@10 on \beir-13. Results for sparse baselines are taken from the original papers, which typically report only on this subset of \beir. Table~\ref{tab:offshelf} reports Beir-15 results for \sparseup models against \textsc{DenseOn}, \textsc{LateOn}, and \textsc{SlideOn}.}
\label{tab:beir13}
\setlength{\tabcolsep}{3.5pt}
\resizebox{\textwidth}{!}{%
\begin{NiceTabular}{lrrrrrrrrrrrrrr}
\CodeBefore
  \tikz \fill[linkup-green!25, rounded corners=3pt] (14-|1) rectangle (15-|last);
  \tikz \fill[linkup-green!70, rounded corners=2pt]
        ([xshift=1pt,yshift=-1pt]14-|7) rectangle ([xshift=-1pt,yshift=1pt]15-|8);
  \tikz \fill[linkup-green!70, rounded corners=2pt]
        ([xshift=1pt,yshift=-1pt]14-|15) rectangle ([xshift=-1pt,yshift=1pt]15-|16);
\Body
\toprule
Model & Arg & Cli & DBP & FEV & FiQ & Hot & NFC & NQ & Quo & SciD & SciF & TRE & Tou & Avg  \\
\midrule
\multicolumn{8}{l}{\textsc{Competitive sparse baselines}} \\
\spladeT & 50.9 & 23.3 & 45.0 & 79.6 & 37.4 & 69.2 & 35.7 & 58.6 & 81.4 & 15.8 & 71.0 & 74.8 & 29.3 & 51.7  \\
\textsc{Granite-30M} & 54.7 & 25.5 & 37.2 & 82.0 & 35.2 & 68.1 & 32.3 & 50.8 & 85.1 & 21.3 & 71.4 & 71.9 & 22.8 & 50.6  \\
\textsc{VT} & 49.0 & 27.8 & 45.1 & 84.4 & 36.3 & 68.8 & 35.8 & 55.9 & 83.6 & 15.6 & 70.5 & 75.9 & 33.0 & 52.4  \\
\textsc{OS-NSE} v1 & 50.8 & 26.3 & 44.7 & 82.1 & 37.6 & 69.7 & 36.0 & 55.3 & 85.6 & 16.4 & 72.3 & 77.1 & 27.8 & 52.4  \\
\textsc{OS-NSE doc-v3} & 52.0 & 31.2 & 45.5 & 86.0 & 40.7 & 71.6 & 36.0 & 58.2 & 87.3 & 16.7 & 72.5 & 73.4 & \textbf{38.9} & 54.6  \\
\laconic (1B) & \textbf{62.1} & 37.4 & 46.8 & 88.3 & 43.4 & 79.2 & \textbf{39.2} & \textbf{66.3} & 85.2 & \textbf{24.2} & 75.6 & \textbf{83.9} & 31.2 & 58.7  \\
\midrule
\multicolumn{8}{l}{\textsc{Comparable training setting}} \\
\denseon & 54.7 & 37.5 & 44.7 & 90.7 & \textbf{53.9} & 74.5 & 39.0 & 59.3 & 89.3 & 22.4 & 76.0 & 82.3 & 28.4 & 57.9 \\
\lateon & 50.5 & 39.7 & 46.0 & \textbf{92.0} & 53.1 & \textbf{80.0} & 37.8 & 63.9 & \textbf{89.7} & 21.9 & \textbf{76.6} & 83.6 & 30.5 & 58.9  \\
\textsc{SlideOn} & 47.2 & \textbf{43.9} & \textbf{48.5} & \textbf{92.0} & 49.7 & 79.8 & 38.6 & 63.9 & 86.0 & 20.2 & 75.7 & 79.2 & 37.9 &  58.7 \\
\midrule
\sparseup & 56.8 & 36.1 & 41.5 & 89.7 & 46.9 & 77.2 & 35.1 & 57.1 & 86.6 & 18.4 & 73.9 & 78.4 & 35.0 & 56.4 \\
\sparseupvtwo & 56.1 & 41.1 & 47.6 & 91.5 & 50.9 & \textbf{80.0} & 38.9 & 63.6 & 87.1 & 20.3 & 75.1 & 79.2 & 38.0 & \textbf{59.2} \\
\end{NiceTabular}}
\end{table}

\section{\sparseupvtwo}\label{sec:sparseup-v2}

Building on the findings of Section~\ref{sec:experiments}, we extend the \sparseup recipe with four components: large-scale contrastive pre-training, cross-encoder distillation, Matryoshka-style training, and long-context training.
\subsection{Contrastive Pre-Training}\label{sec:cpt}

The \denseon and \lateon recipe~\citep{sourty2026denseonlateonfullyopen} comprises two retrieval-oriented training stages: \emph{(i)}~a large-scale contrastive pre-training (CPT) stage on 665M weakly supervised pairs, most of which use document titles as queries and the corresponding passages as positives; and \emph{(ii)}~a supervised fine-tuning stage on 1.88M pairs drawn from standard collections such as TriviaQA and \msmarco, which provide relevance labels along with teacher scores from a strong cross-encoder.

\begin{figure}[t]
    \centering
    \includegraphics[width=1.0\linewidth]{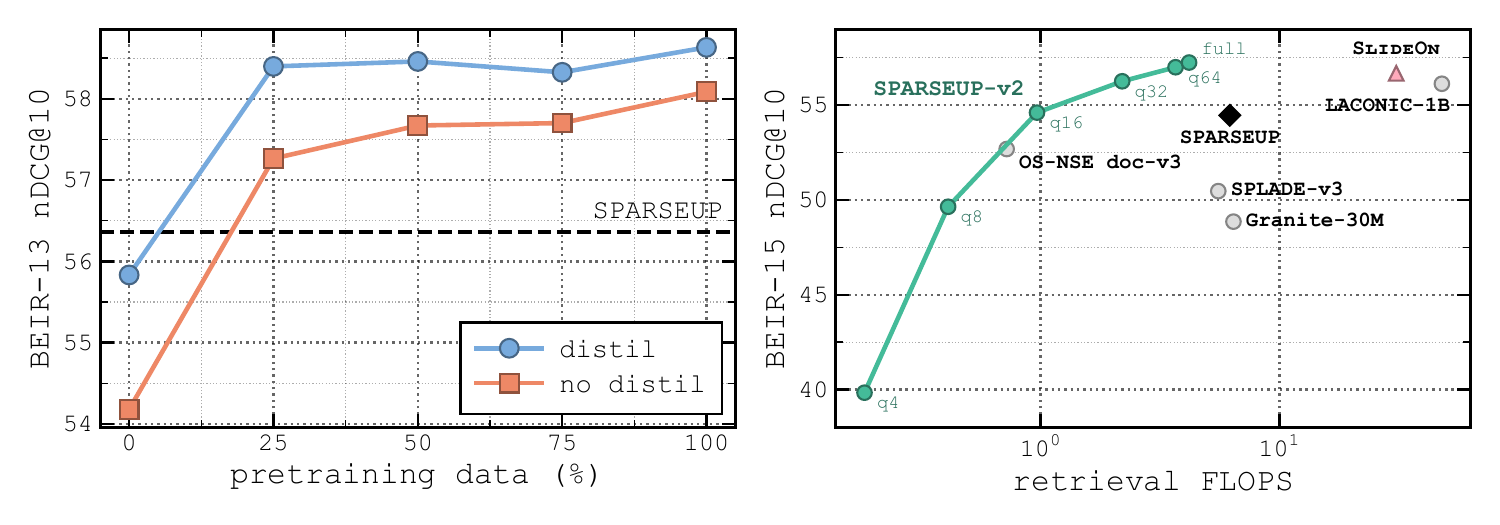}
    \caption{\textit{(Left)} \beir-13 nDCG@10 as a function of the fraction of contrastive pre-training (CPT) data used, with CPT performed directly in the sparse setting, followed by fine-tuning with and without distillation. The leftmost point (0) corresponds to \modernbert without CPT. The dashed line marks \sparseup, initialized from the \textsc{LateOn-unsupervised} checkpoint (Section~\ref{sec:experiments}). Queries and documents are truncated to 64 and 512 tokens, respectively. \textit{(Right)} \beir-15 nDCG@10 as a function of retrieval FLOPS. The \sparseupvtwo curve covers query budgets of $\{4, 8, 16, 32, 64\}$ non-zero entries, with documents capped at 256 non-zero entries, as well as the unpruned model. {\bf \texttt Across all budgets, \sparseupvtwo lies on the Pareto front, offering the best effectiveness for a given retrieval cost among the compared models}.}
    \label{fig:cpt}
\end{figure}

In Section~\ref{sec:experiments}, we skipped stage \emph{(i)} by initializing \sparseup from the \textsc{LateOn-unsupervised} CPT checkpoint, which was pre-trained in the \emph{late interaction} setting. This initialization provides a clear advantage over starting from the bare \modernbert backbone (Figure~\ref{fig:cpt}, {\it Left}: dashed line for \sparseup \emph{vs} leftmost square for \modernbert). However, it leaves a mismatch between pre-training and fine-tuning: the backbone was never optimized to produce sparse lexical representations, and the MLM head was not trained jointly with the contrastively pre-trained encoder. Such a mismatch has been shown to degrade downstream performance~\citep{chaffin2026colbertzeropretrainpretraincolbert}. We therefore run contrastive pre-training with the sparse architecture on increasing fractions of the pre-training corpus, and then apply the same supervised stage as \sparseup, with and without distillation. We follow a setup similar to that of \denseon and \lateon, using large batches of 16$k$ pairs with in-batch negatives, made feasible by GradCache~\citep{gao2021scalingdeepcontrastivelearning}.

Figure~\ref{fig:cpt} reports \beir-13 nDCG@10 as a function of the fraction of CPT data used; the \textsc{``0''} point corresponds to the \modernbert backbone without CPT. The figure displays the clear benefit of a dedicated CPT. With only $25\%$ of the data, sparse CPT improves over the \modernbert initialization by roughly $3$ points. It also outperforms \sparseup initialized from the \lateon checkpoint by roughly $1$ point without distillation, and $2$ points with it. Cross-modality pre-training is therefore better than no pre-training, but it consistently lags behind pre-training in the target modality. This advantage comes at a price: the full CPT run takes roughly $3.5$ days on $4$ NVIDIA H100 GPUs, which is $50\times$ the time it takes for fine-tuning. We also release the model after weakly supervised pre-training to ease the development of competitive sparse encoders.

\subsection{Knowledge Distillation}

We evaluate the effect of distillation on top of the \sparseup recipe. We reuse the \data training data from the supervised stage, which also provides teacher scores from a strong cross-encoder reranker, \texttt{mxbai-rerank-large-v2}~\cite{li2025prorank}. As in \textsc{SlideOn}~\citep{moura2026slideon}, we extend the original fine-tuning mixture with the English training splits of MIRACL~\citep{zhang-etal-2023-miracl} and MLDR~\citep{bge-m3}, the latter improving the ability of \sparseupvtwo to handle long-document retrieval.
We adopt the training objective of the \textsc{mDenseOn} and \textsc{mLateOn} recipes~\citep{sourty2026denseonlateonfullyopen}, which combines the InfoNCE loss with a Kullback--Leibler divergence term, weighted by $\lambda_w$, between the teacher and student score distributions over the candidate documents.

We report the results in Figure~\ref{fig:cpt}, evaluating knowledge distillation together with contrastive pre-training, since their effects are intertwined. Distillation helps at every CPT budget, but its benefit shrinks as the backbone sees more CPT data: the gap between the distilled and non-distilled models goes from $+1.7$ points without CPT to $+1.1$ at 25\%, $+0.8$ at 50\%, $+0.6$ at 75\%, and $+0.5$ at 100\%. As a result, a distilled model trained on only 25\% of the CPT data (58.4) already slightly outperforms a non-distilled model trained on all of it (58.1). Since the cost of fine-tuning is negligible compared to that of pre-training, distillation acts as a cheap substitute for a large share of the pre-training budget. Finally, combining full CPT and knowledge distillation reaches 58.6, $+2.2$ points over \sparseup{}.

\subsection{Matryoshka Learning}\label{sec:matryoshka}

Learned sparse representations can be pruned at inference time with limited loss in effectiveness~\citep{lassance2023static-pruning}. This allows a single model to serve multiple effectiveness--efficiency trade-offs without retraining, simply by choosing how many entries to keep per query and per document. In Section~\ref{sec::ablation}, we showed that training at a fixed top-$k$ does not provide this property: the model improves at the training budget, but degrades when more entries are kept at inference time. As the last ingredient of \sparseupvtwo, we therefore adopt Matryoshka-style training~\citep{kusupati2022matryoshka}, which optimizes a single model for several pruning budgets at once, and has been shown to work for dense and late interaction approaches~\cite{sourty2026denseonlateonfullyopen}.

Let $q, d \in \mathbb{R}^{|\mathcal{V}|}$ be the sparse representations of a query and a document, and let $\text{top}(x, k)$ denote the operator that keeps the $k$ largest entries of $x$ and sets all others to zero. We define $m$ levels $\{(a_i, b_i)\}_{i=1}^{m}$, where $a_i$ and $b_i$ are the number of entries kept for queries and documents, respectively. Let $\mathcal{L}(q, d)$ be the training loss of the previous sections (InfoNCE plus the KL distillation term). The Matryoshka loss is a weighted sum of this loss over all levels:
\begin{equation}
    \mathcal{L}_{\text{mat}} = \sum_{i=1}^{m} \alpha_i \, \mathcal{L}\big(\text{top}(q, a_i),\, \text{top}(d, b_i)\big),
\end{equation}
where $\alpha_i$ is the weight of level $i$. We use the levels $\{(16, 64), (32, 128), (64, 256)\}$, plus the unpruned representations as a fourth level ($a_4 = b_4 = |\mathcal{V}|$), with weights $\alpha = \{0.05, 0.08, 0.1, 0.8\}$, which gave the best trade-off between pruned and unpruned effectiveness. The FLOPS regularizer is applied only once, on the full representations.

\begin{figure}[t]
    \centering
    \includegraphics[width=1.0\linewidth]{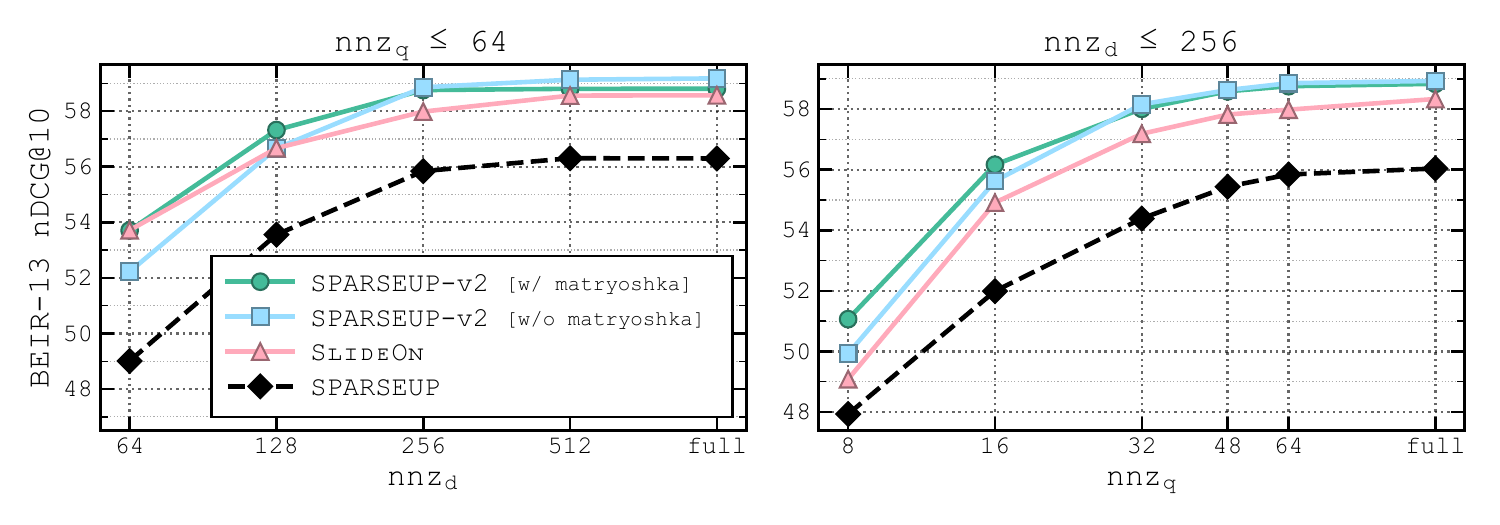}
    \caption{\beir-13 nDCG@10 under inference-time pruning for \sparseupvtwo trained with and without Matryoshka, \textsc{SlideOn}, and \sparseup. {\it (Left)} Queries capped at $64$ non-zero entries, varying the document cap. {\it (Right)} Documents capped at $256$ non-zero entries, varying the query cap. ``full'' denotes no cap.}
    \label{fig:matryoshka}
\end{figure}

Figure~\ref{fig:matryoshka} reports \beir-13 nDCG@10 at different query and document pruning budgets. Matryoshka training pays off where pruning is most aggressive. With queries capped at $64$ entries {\it (Left)}, it improves nDCG@10 by $1.5$ points at $64$ entries per document and by $0.7$ points at $128$. With documents capped at $256$ entries {\it (Right)}, it improves nDCG@10 by $1.1$ points at $8$ entries per query and by $0.5$ points at $16$. At larger budgets the two variants are on par, and without pruning Matryoshka costs at most $0.4$ points. The Matryoshka model also outperforms \sparseup by $2.5$ to $4.7$ points at every budget. It matches \textsc{SlideOn} at the most aggressive document cap and outperforms it everywhere else, with the largest margins under aggressive query pruning ($+2.0$ points at $8$ entries per query and $+1.3$ at $16$).

\subsection{Comparison with the State of the Art}

Table~\ref{tab:beir13} compares our models with several state-of-the-art sparse encoders: \spladeT~\citep{lassance2024spladev3}, \textsc{Granite-30M-sparse}~\citep{awasthy2025graniteembeddingmodels}, \modernbert with Vocabulary Transfer~\citep{DBLP:conf/sigir/GengY26} (\textsc{VT}), OpenSearch Neural Sparse Encoding~\citep{10.1145/3726302.3730192} (\textsc{OS-NSE v1} and \textsc{OS-NSE doc-v3}) and \laconic~\citep{xu2026laconicdenseleveleffectivenessscalable}. We also include \denseon, \lateon~\citep{sourty2026denseonlateonfullyopen}, and \textsc{SlideOn}~\cite{moura2026slideon} to compare our sparse retriever against dense, late interaction and latent sparse encoders trained on the same dataset and backbone. 

\sparseup reaches an average nDCG@10 of 56.4 on \beir-13, lagging behind both \denseon and \lateon. 
Replacing the \lateon initialization with our sparse contrastive pre-training raises it to $58.1$ (Figure~\ref{fig:cpt}, {\it left}, rightmost orange square). This configuration shares the training stages and data of \denseon and \lateon, and the sparse model lands between the two: above \denseon ($57.9$) and below \lateon ($58.9$). 

\sparseupvtwo is instead directly comparable with \textsc{SlideOn}, since both also include MIRACL-en~\cite{zhang-etal-2023-miracl} in their fine-tuning data: it outperforms \textsc{SlideOn} by $0.5$ nDCG@10 points on \beir-13 ($59.2$ \emph{vs} $58.7$), showing the effectiveness of vocabulary-based sparse models. Note that \sparseupvtwo outperforms \textsc{Laconic-1B}, which has roughly an order of magnitude more parameters (and non-zero entries), by the same margin. 

Table~\ref{tab:offshelf} extends the comparison with \denseon, \lateon, and \textsc{SlideOn} to a broader set of specialized benchmarks: CURE-en~\cite{10.1145/3711896.3737435} for medical retrieval; LegalPinCite~\cite{rampisela-etal-2026-legalpincite} for paragraph-level legal retrieval over case law from the Court of Justice of the European Union (CJEU); Finance, the average over the FinQA and FinanceBench tasks of MTEB~\cite{muennighoff-etal-2023-mteb}; ChemTEB~\cite{pmlr-v262-shiraee-kasmaee24a} for chemistry; BRIGHT-Pro~\cite{zhao-etal-2026-rethinking} for reasoning-intensive retrieval; LoTTE~\cite{Santhanam2022ColBERTv2} and FreshStack~\cite{thakur2025freshstack} for long-tail and technical question answering; LIMIT~\cite{weller2025theoretical}, which probes the representational limits of embedding-based retrieval; the English subsets of MIRACL~\cite{zhang-etal-2023-miracl} and MLDR~\cite{bge-m3}, the latter targeting long-document retrieval; and \decbeir~\cite{sourty2026denseonlateonfullyopen}, a decontaminated variant of \beir. \sparseupvtwo obtains the best nDCG@10 on 5 of the 13 benchmarks (\beir-15, \textsc{Decon-Beir}, CURE-en, Finance, and BRIGHT-Pro), with the largest gains on reasoning-intensive and domain-specific retrieval: $+2.0$ nDCG@10 over the best baseline on BRIGHT-Pro and $+4.1$ on Finance. Compared with \denseon, trained under a closely comparable pipeline, \sparseupvtwo is better on 10 of 13 benchmarks, with particularly large gaps on MLDR-en ($+20.2$) and LIMIT, where the dense model nearly fails (0.8 \emph{vs} 28.2 R@2). \lateon remains the strongest model on long-document retrieval (MLDR-en), on MIRACL-en, and especially on LIMIT, and is slightly ahead on LoTTE and FreshStack ($\leq 1.8$ points on the primary metric). Finally, compared with the SAE-based \textsc{SlideOn}, \sparseupvtwo is better on 11 of 13 benchmarks, the exceptions being ChemTEB, where \textsc{SlideOn} leads by 4.3 points, and MIRACL-en, where they are on par (60.7 \emph{vs} 60.9).

Relative to \sparseup, the full pipeline improves the primary metric on 11 of 13 benchmarks, most notably on BRIGHT-Pro ($+14.9$), LIMIT ($+20.1$ R@2), and MLDR-en ($+4.0$). The only regressions are on LegalpinCite-par ($-2.9$ nDCG@10) and, marginally, on FreshStack ($-0.5$ $\alpha$-nDCG@10).

\begin{table}[t]
\centering
\caption{Full evaluation suite. CURE-en: mean of the 10 English CUREv1 splits. Finance: mean of FinQA and FinanceBench (MTEB). LegalPincite-par~\cite{rampisela-etal-2026-legalpincite}: the paragraph-level retrieval on CJEU data. ChemTEB: mean of ChemHotpotQA and ChemNQ. BRIGHT-Pro: mean of 7 domains. \decbeir: Decontaminated BEIR.}
\label{tab:offshelf}
\small
\setlength{\tabcolsep}{4pt}
\begin{NiceTabular}{ll rrr rr}
\CodeBefore
  \lknumband{2}{29}{7}
  \lkcolbest{2}{7}\lkcolbest{3}{7}\lkcolbest{4}{7}
  \lkcolbest{6}{7}\lkcolbest{7}{7}
  \lkcolbest{9}{7}\lkcolbest{10}{7}\lkcolbest{11}{7}
  \lkcolbest{14}{7}\lkcolbest{15}{7}
  \lkcolbest{22}{7}
\Body
\toprule
Dataset & Metric & \textsc{DenseOn} & \textsc{LateOn} & \textsc{SlideOn} & \sparseup & \sparseupvtwo\\
\midrule
                               
\multirow{2}{*}{BEIR-15}    & nDCG@10 & 56.2 & \textbf{57.2 }& 56.7 & 54.5 & \textbf{57.2} \\
                            & R@100   & 74.0 & 74.2 & 74.7 & 71.5 & \textbf{74.8} \\

\midrule
\multirow{2}{*}{\decbeir} & nDCG@10 & 57.9 & 59.1 & 58.4 & 57.5 & \textbf{59.5} \\
                          & R@100   & 73.3 & \textbf{74.2} & 73.3 & 72.1 & 74.0 \\
                               \midrule
\multirow{2}{*}{CURE-en}       & nDCG@10 & 56.5 & 61.3 & 60.8 & 56.0 &  \textbf{62.1} \\
                               & R@100   & 75.6 & 77.8 & 77.9 & 73.1 &  \textbf{78.1} \\
\midrule
\multirow{2}{*}{LegalPinCite-par} & nDCG@10 & 46.3 & 54.7 & 48.2 & \textbf{55.2} &  52.3 \\
                               & R@100   & 77.2 & 79.9 & 77.0 & \textbf{81.5} &  \textbf{81.5} \\
\midrule
\multirow{2}{*}{Finance} & nDCG@10 & 59.8 & 59.2 & 62.6 & 63.4 &  \textbf{66.7 }\\
                               & R@100   & 96.0 & 92.6 & 96.5 & \textbf{96.9} &  \textbf{96.9} \\
\midrule
\multirow{2}{*}{ChemTEB}       & nDCG@10 & 76.4 & 75.1 & \textbf{76.8} & 70.8 &  72.5 \\
                               & R@100   & 96.9 & 95.4 & 96.9 &\textbf{ 97.5}  & 96.9 \\
\midrule
\multirow{2}{*}{BRIGHT-Pro}    & nDCG@10 & 29.6 & 36.1 & 47.4 & 34.5  & \textbf{49.4} \\
                               & R@100   & 62.9 & 70.7 & 77.6 & 68.7  & \textbf{80.0} \\
\midrule
\multirow{2}{*}{LoTTE Forum}   & S@5     & \textbf{82.7} & 81.5 & 78.0 & 75.5 & 80.6 \\
                               & R@100   & \textbf{78.1} & 75.4 & 72.3 & 67.9  & 73.3 \\
\midrule
\multirow{2}{*}{LoTTE Search}  & S@5     & 83.4 & \textbf{83.9} & 81.1 & 78.2  & 82.1 \\
                               & R@100   & \textbf{88.4} & 87.3 & 85.9 & 83.8  & 86.5 \\
\midrule
\multirow{3}{*}{FreshStack}    & $\alpha$-nDCG@10 & 19.6 & \textbf{24.5} & 15.6 & 23.3 & 22.8 \\
                               & C@20    & 45.4 & \textbf{51.7} & 40.1 & 51.0 &  51.6 \\
                               & R@50    & 31.1 & 35.5 & 28.3 & 35.3 &  
                               \textbf{36.3} \\
\midrule
\multirow{2}{*}{MIRACL-en}    & nDCG@10 & 55.4 & \textbf{64.3} & 60.9 & 56.7 & 60.7 \\
                               
                               & R@100    & 96.6 & 96.6  & \textbf{97.7} & 96.4 &  97.5 \\
\midrule
\multirow{2}{*}{MLDR-en}  & nDCG@10 & 41.7 & \textbf{71.6} & 58.7 & 57.9 & 61.9 \\
                          & R@100   & 65.8 & \textbf{92.6} & 85.4 & 84.8 & 87.3 \\
\midrule
\multirow{3}{*}{LIMIT} & R@2   & 0.8 & \textbf{74.2} & 12.2 & 8.1  & 28.2 \\
                       & R@10  & 2.1 & \textbf{89.7} & 23.5 & 16.2 & 48.5 \\
                       & R@100 & 6.8 & \textbf{96.2} & 48.3 & 34.7 & 74.1 \\

\bottomrule
\end{NiceTabular}
\end{table}

\section{Concluding Remarks}

We presented a recipe for training \splade-like encoders on modern backbones, turning architectures that underperform or fail under the standard recipe into effective sparse retrievers. Its key ingredients, logit shifting, fan-out control, and vocabulary folding, each improve the effectiveness--efficiency trade-off of sparse encoders, and their benefits combine. Scaled with contrastive pre-training, distillation, and Matryoshka-style training, our \sparseupvtwo model reaches 57.2 average nDCG@10 on \beir-15, matching \lateon, the best public retriever under 150M parameters we know of. Because \denseon, \lateon, and \sparseupvtwo share closely comparable training pipelines, our results also provide a controlled comparison across paradigms: out of domain, learned sparse representations sit between dense and late-interaction models in effectiveness, at a fraction of the latter's cost. We hope these findings renew interest in sparse retrieval, both as a standalone first-stage retriever and as an efficient candidate generator for multi-stage pipelines.
\bibliography{references}
\bibliographystyle{plainnat}

\beginappendix

\end{document}